\documentstyle[12pt]{article}

\newcommand{\be}{\begin{equation}}
\newcommand{\ee}{\end{equation}}
\newcommand{\bn}{\begin{eqnarray}}
\newcommand{\en}{\end{eqnarray}}
\newcommand{\p}{\partial}

\def\ni{\noindent}
\input{tcilatex}

\begin{document}

\begin{center}
{\Large {\bf The non-Abelian, PST and supersymmetric formulations of Hull's notons} } 
\vspace{6mm}

\noindent{\ 
{\large Everton M. C. Abreu}\footnote{Present address: Depto. de Campos e Part\'{\i}culas, Centro Brasileiro de 
Pesquisas F\'{\i}sicas, Rua Xavier Sigaud 150, Urca, 22290-180, 
Rio de Janeiro, Brazil. {\sf E-mail: everton@cbpf.br}} 
}\vspace{3mm}

\noindent 
{\large 
{\it The Abdus Salam International Centre for Theoretical Physiscs \\[0pt]
Strada Costiera 11, 34014, Trieste, Italy\\[0pt]
and \\[0pt]
Instituto de Ci\^encias, Universidade Federal de Itajub\'a, \\[0pt]
Av. BPS 1303, Pinheirinho, Itajub\'a, \\[0pt]
Minas Gerais, Brazil} } 
\vspace{4mm}

\today

\vspace{14mm}
\end{center}

\begin{abstract}
\noindent Chiral $p$-forms are, in fact, present in many supersymmetric and
supergravity models in two, six and ten dimensions. In this work, the dual
projection procedure, which is essentially equivalent to a canonical
transformation, is used to diagonalize some theories in $D=2$ ($0$-forms).
The dual projection performed here provides an alternative way of gauging
the chiral components without the necessity of constraints. 
It is shown that the nonmover field (the noton) initially
introduced by Hull to cancel out the Siegel anomaly, has non-Abelian, PST
and supersymmetric formulations.
\end{abstract}

\newpage

\section{Introduction}

Chiral $p$-forms enter in the spectrum of type IIB superstring and it can be
shown that a string can couple directly to a chiral $p$-form \cite{schwarz}.
A chiral $p$-form has a number of properties which are very peculiar.
Although its excitations obey Bose statistics, it shares many of the
features of fermionic fields like a field equation which is linear in
derivatives; it leads to species doubling if one attempts to define it on a
lattice \cite{tang} and it gives rise to gravitational anomalies \cite{awebn}%
.

We can describe chiral bosons by a $p$-form gauge potentials $B_p$. The $B_p$%
's curvatures $H_{p+1}=d B_p$ satisfy, as equations of motions, a Hodge
(anti) self-duality condition in a space-time with dimension $D=2(p+1)$. The
values of $p$ are restricted to even values by the self-consistency of such
equation in space-times with Minkowskian signature $\eta_{ab} =
(1,-1,-1,\ldots,-1)$. The restriction on $p$ makes $D=2,6,10,\ldots$, be the
relevant dimensions of the chiral boson theories.

Chiral bosons are important in superstring and supergravity theories, and
more recently $M$ theory. Two dimensional chiral bosons (scalar) are basic
ingredients in string theory. The six-dimensional ones belong to the
supergravity and tensor multiplets in $N=1, D=6$ supergravity theories. They
are necessary to complete the $N=2, D=6$ supermultiplet of the $M$-theory
five-brane. Finally a ten dimensional chiral bosons appears in $IIB, D=10$
supergravity.

In this work we study the dual projection technique introduced by Wotzasek 
{\it et al} in \cite{wotzasek,baw,baw2}. This technique, strictly related to
canonical transformation \cite{bg}, has been presented in the study of
electromagnetic duality groups. The difference between both concepts is that 
the dual projection
is performed at the level of the actions while the canonical transformation
is at the Hamiltonian level. However, in this work we will consider only
first-order actions. Furthermore, the equivalence between the dual
projection and the canonical transformation becomes manifest \cite{bg}. The
most useful and interesting point in this dual projection procedure is that
it is not based on evidently even-dimensional concepts and may be extended
to the odd-dimensional situation. But we will not analyze this feature here.

This work is motivated by the fact that coupling chiral fields to external
gravitational fields reveals the presence of notons \cite{hull,dgr}. Noton
is a nonmover field at classical level, carrying the representation of the
Siegel symmetry \cite{siegel}, that acquires dynamics upon quantization. At
the quantum level, it was shown \cite{aw}, that its dynamics is fully
responsible for the Siegel anomaly.

The Wess-Zumino-Witten-Novikov (WZWN) \cite{witten,novikov} model is a conformal field theory that has been used in the past to reproduce various two dimensional systems like Toda
field theories, black holes and others. Recently it was carried out the
connection of this model to a combination of three dimensional topological
BF and Chern-Simons gauge theories defined on a manifold with boundaries
using a direct application of non-Abelian T duality on the WZWN nonlinear
sigma model \cite{mohammed}. 

In this work we discuss the dual projection of
non-Abelian chiral fields coupled to external gravitational backgrounds
which is a result from the diagonalization of the first-order form of an action that
describes the chiral WZWN model.

The Lorentz covariant approach to describe chiral bosons, proposed by Pasti,
Sorokin and Tonin (PST) \cite{pst1}, has been studied in \cite
{lechner,blnpst,lechner2}. The quantum behavior of the PST models in the
framework of the algebraic renormalization \cite{ps} was analyzed in \cite
{dps}. Differently of \cite{varios} it introduces only one scalar auxiliary
field, but in a non-polynomial way.

The PST procedure has important roles on the construction of a covariant
effective action for the $M$-theory five-brane \cite{blnpst}, covariant
actions for supersymmetric chiral bosons \cite{lechner2} and to rederive the
gravitational anomaly for the chiral bosons \cite{lechner}.

Here we disclose the presence of the noton in four models using the dual
projection. In one we corroborate, in a clearer manner, the result found in 
\cite{wotzasek2}. Hence, we propose three new formulations for the Hull
noton: two different Lorentz invariant versions of PST Siegel's chiral boson 
and a supersymmetric formulation.

The sequence of the paper is: in the next section we review the dual
projection applied to the Abelian chiral model and to the Principal Chiral
Model as examples. In section III we carry out the dual projection of the
chiral WZWN model. The analysis of two different versions of PST $D=2$
chiral bosons and the supersymmetric one (\cite{pst2}) are analyzed in section IV.
Finally, the conclusion and perspectives are discussed in the last section.

\section{The dual projection: a review}

The separation of a scalar field into its chiral components has been
introduced by Mandelstam \cite{mandelstam} in its seminal paper on 2D
bosonization. The chiral splitting for the non-Abelian side has been studied
by Polyakov and Wiegman \cite{pw}. However, the Abelian limit of
Polyakov-Wiegman decomposition does not coincide with Mandelstam's chiral
decomposition. The chiral separation in \cite{mandelstam} is based on a
first-order theory while that of \cite{pw} is second-order. Mandelstam's
chiral decomposition scheme can be obtained as a constraint over the theory
requiring the complete separability of the original action.

The dual projection in $D=2(2p+1)$ dimensions leads to a diagonal form of
the action. Here, differently from \cite{bw}, the two pieces manifest
completely unlike features: while one piece is chiral, responsible for the
dynamic sector of the theory, the other carry the algebraic component, as
shown in \cite{aw}. We select here only two examples, among others, in the
literature where the dual projection succeed.

\subsection{The Siegel chiral boson}

In this section we will make a brief review of the main results obtained in 
\cite{aw}. We begin with the Siegel original classical Lagrangian density
for a chiral scalar field \cite{siegel}, 
\begin{equation}  \label{um}
{\cal L}\,=\,\partial_+\phi\,\partial_-\phi\,-\,\lambda_{++}\,(\partial_-%
\phi)^2\,=\,{\frac{1 }{2}}\,\sqrt{-g}\,
g^{\alpha\beta}\,\partial_{\alpha}\,\phi\,\partial_{\beta}\,\phi
\end{equation}
where the metric 
\begin{equation}  \label{dois}
g^{++}\,=\,0\:,\:\:\:\:\:\:g^{+-}\,=\,1\:,\:\:\:\:\:\:g^{--}\,=\,-\,2\,%
\lambda_{++}\:,
\end{equation}
and the Lagrangian (\ref{um}) describes a lefton \cite{gs}. A lefton (or a
righton) is a particle which, besides to carry the dynamics of the theory,
it is liable for the symmetry too. This characterizes exactly a Siegel mode.
It is different from a Floreanini-Jackiw's (FJ's) mode \cite{fj}. Even if
the FJ particle be a left-mover, it is only responsible for the dynamics.
Hence it is not a lefton.

The symmetry content of the theory is well described by the Siegel algebra,
a truncate diffeomorphism that disappears at the quantum level. Hence (\ref
{um}) is invariant under Siegel gauge symmetry which is an invariance under
the combined coordinate transformation and a Weyl rescaling of the form 
\begin{equation}  \label{tres}
x^- \rightarrow \tilde{x}^-\,=\,x^-\,-\,\epsilon^- \quad \mbox{and} \quad
\delta\,g_{\alpha\beta} \,=\, -\,g_{\alpha\beta}\,\partial_-\,\epsilon^-\;\;.
\end{equation}
The fields $\phi$ and $\lambda_{++}$ transform under (\ref{tres}) as
follows: 
\begin{eqnarray}
\delta \phi&=& \epsilon^-\,\partial_-\,\phi  \nonumber \\
\delta \lambda_{++} \,=\,
-\,\partial_+\,\epsilon^-\,&+&\,\epsilon^-\,\partial_-\,\lambda_{++}\,-\,
\partial_-\,\epsilon^-\,\lambda_{++}\;\;,
\end{eqnarray}
and $\phi$ is invariant under the global axial transformation 
\begin{equation}
\phi \rightarrow \tilde{\phi}\,=\,\phi\,+\,\bar{\phi}\;\;.
\end{equation}
It is well known that fixing the value of the multiplier as $\lambda_{++}=1$
in (\ref{um}) we can obtain the FJ model.

We will now begin to introduce the dual projection. This is done introducing
a dynamical redefinition in the phase space of the model. Using (\ref{um})
in Lorentz coordinates we can obtain the canonical momentum as 
\begin{equation}  \label{quatro}
\pi\,=\,\frac{\partial {\cal L}}{\partial \dot{\phi}}\,=\,\dot{\phi}%
\,-\,\lambda_{++}\,(\,\dot{\phi}\, -\,\phi^{\prime}\,)
\end{equation}
and so we have that 
\begin{equation}  \label{cinco}
\dot{\phi}\,=\,\frac{\pi\,-\,\lambda_{++}\,\phi^{\prime}}{1\,-\,\lambda_{++}}%
\;\;.
\end{equation}
After a little algebra, substituting (\ref{quatro}) and (\ref{cinco}) into (%
\ref{um}), the Lagrangian in the first-order form reads 
\begin{equation}  \label{cincoa}
{\cal L}\,=\,\pi\,\dot{\phi}\,-\,\frac{{\phi^{\prime}}^2}{2}\,-\, {\frac{1 }{%
2}}\,\frac{(\,\pi\,-\,\lambda_{++}\,\phi^{\prime}\,)^2}{1\,-\,\lambda_{++}}%
\,-\, {\frac{\lambda_{++} }{2}}\,{\phi^{\prime}}^2\;\;.
\end{equation}

As we said above, we have to fix the value of the multiplier as $\lambda
_{++}\rightarrow 1$ to get the FJ form. This value of $\lambda _{++}$
promotes a reduction of the phase space of the model to \cite{djt} 
\begin{equation}
\label{novepi}
\pi \rightarrow {\phi ^{\prime }}\;\;,  \label{seis}
\end{equation}
and consequently the third term in (\ref{cincoa}) reduces to zero as $%
\lambda _{++}\rightarrow 1$. Therefore the dynamics of the system will be
described by a FJ action. The above behavior in (\ref{novepi}) suggests the following
canonical transformations: 
\begin{equation}
\phi \,=\,\varphi \,+\,\sigma \qquad \mbox{and}\qquad \pi \,=\,\varphi
^{\prime }\,-\,\sigma ^{\prime }\;\;,  \label{sete}
\end{equation}
and we stress that these fields are independent as they originate from
completely different actions. After substituting (\ref{sete}) into (\ref
{cincoa}) to perform the dual projection we find a diagonalized Lagrangian, 
\begin{equation}
{\cal L}\,=\,\varphi ^{\prime }\,\dot{\varphi}\,-\,{\varphi ^{\prime }}%
^{2}\,-\,\sigma ^{\prime }\,\dot{\sigma}\,-\,\eta _{+}\,{\sigma ^{\prime }}%
^{2}  \label{oito}
\end{equation}
where 
\[
\eta _{+}\,=\,\frac{1\,+\lambda _{++}}{1\,-\lambda _{++}}\;\;. 
\]
The effect of dual projection procedure into the first-order Siegel theory,
equation (\ref{cincoa}) was the creation of two different internal spaces
leading to the $Z_{2}$ group of dualities (a discrete group with two
elements) \cite{wotzasek,baw} and the other is the diffeomorphism group of
transformations. Clearly we can see that the chirality/duality group and the
symmetry group are in different sectors. The first is obviously a FJ mode
and the other is a noton mode. This result is complementary to the
established knowledge, where the FJ action is interpreted as a gauge fixed
Siegel action \cite{siegel}. Under this point of view, we look at the gauge
fixing process as the condition that sets the noton field to vanish. It can
be proved \cite{aw} that this noton is totally responsible for the
symmetries, both classically and quantically. We finally can say that the
importance of the inclusion of a normalized external noton - the Hull
mechanism - is, conveniently, to cancel the Siegel anomaly \cite{hull}.

\subsection{The Principal Chiral Model}

As a non-Abelian example, let us next extend the separability condition, due
to the dual projection, discussed above to the non-Abelian bosons following
only the main steps given in \cite{bw}. 

The most obvious choice would be to
consider an action given by a bilinear gradient of a matrix-valued field $g$
taking values on some compact Lie group $G$, which would be the natural
extension of the free scalar Abelian field. This is the action for the
Principal Chiral Model (PCM) that reads 
\begin{equation}  \label{pcm}
{\cal S}_{PCM}(g)= {\frac{1}{2}} \int d^2 x \,tr \left( \partial_\mu g
\;\partial^\mu \tilde{g} \right)\;\;.
\end{equation}

\noindent Here $g:R^{1,1}\rightarrow G$ is a map from the 2 dimensional
Minkowski space-time to $G$ and let us write for simplicity $\tilde{g}=g^{-1}$. 
This action, however,
puts some difficulties. First of all, by examining its field equation we
learn that, it does not represent a free field. The Jacobian of the field
redefinition does not involve a time-derivative and can be reabsorbed in the
normalization of the partition function. Let us write the PCM in its
first-order form as 
\begin{equation}  \label{WZW1}
{\cal S}_{PCM}(g,P)\,=\, \int d^2 x\, tr \left( \partial_\tau g\, P \right) \,+\, {\frac{ 1}{2 }} \int d^2 x \,tr \left( PgPg \right) \,+\, {\frac{ 1}{2}} \int
d^2 x\, tr \left( \tilde{g}\partial_\sigma g\, \tilde{g}\partial_\sigma g
\right)
\end{equation}

\noindent and redefine the fields $g$ and $P$, which is an auxiliary field, 
through non-Abelian canonical transformations 
\begin{equation}  \label{nonabemand}
g \,=\, A \, B \qquad \mbox{and} \qquad P \,=\, \varepsilon\left(\tilde{B}%
\partial_\sigma \tilde{A} -\partial_\sigma \tilde{B} \tilde{A}\right)
\end{equation}
where, remembering our notation, we have that: $\tilde{A},\tilde{B}%
=A^{-1},B^{-1}$ respectively. The action for the PCM now reads 
\begin{equation}  \label{pcm2}
{\cal S}_{PCM}\left(A,B\right)\,=\, {\cal S}_{\varepsilon}\left(A\right)+ 
{\cal S}_{-\varepsilon}\left(B\right) \,+\, \varepsilon \int d^2x tr\left[%
\tilde{A} \left(\partial_\tau A \partial_\sigma B - \partial_\sigma A
\partial_\tau B\right)\tilde{B}\right]
\end{equation}

\noindent where 
\begin{equation}
{\cal S}_{\varepsilon}\left(A\right)=\int d^2x tr\left(\varepsilon
\partial_\sigma \tilde{A}\partial_\tau A -\partial_\sigma \tilde{A}%
\partial_\sigma A\right)\;\;,
\end{equation}
and ${\cal S}_{-\varepsilon}\,(B)$ is analogous.  We see that due to the non-Abelian nature of the fields, the cross-term cannot be eliminated, so that a complete separation cannot be
achieved. This fact should be expected. In the canonical approach, we have
for their field equations the pair 
\begin{equation}  \label{hje}
\partial_\tau I \,=\, \partial_\sigma J \qquad \mbox{and} \qquad
\partial_\tau J \,=\, \partial _\sigma I - \left[I,J\right]\;\;,
\end{equation}

\noindent where $I= \partial_\tau g\;\tilde{g}$ and $J=\partial_\sigma g\;%
\tilde{g} $. The second equation is a sort of Bianchi identity, which is the
integrability condition for the existence of $g$. Looking at this pair of
equations one can appreciate that chirality is not well defined in this
model. However, this picture changes drastically with the inclusion of the
Wess-Zumino topological term, i.e., when we consider the WZWN model \cite
{witten,novikov}. The first equation in (\ref{hje}) changes to $%
\partial_\tau I = \partial_\sigma J + \rho \left[I,J\right]$, with $\rho \in
Z$, producing a more symmetric set of equations. In particular for $%
\rho=\pm 1$ it is known that the currents above describe two independent
affine Lie algebras. One expects then that with the introduction of the
topological term, one would be able to obtain an identical mixing term, such
that in the total action they could cancel each other. With the topological term 
\begin{equation}
\Gamma_{WZ}(g)={\frac{1}{3}}\int d^3 x\,\epsilon^{ijk}\,tr\left[\,\tilde{g}%
\partial_i g \;\tilde{g}\partial_j g \;\tilde{g}\partial_k g \,\right]\;\;,
\end{equation}
we have now the complete WZWN model defined on the group manifold ${\cal M}%
_G$, based on the Lie algebra $G$, where $g \in {\cal M}_G$ and ${\cal M}$
is a three dimensional ball whose the boundary is the two dimensional
surface $\partial {\cal M}$ \cite{mohammed}.

Using the first of the field redefinitions (\ref{nonabemand}), it is a
lengthy but otherwise straightforward algebra to show that 
\begin{equation}  \label{topol}
\Gamma_{WZ}(A,B) \,=\,\Gamma_{WZ}(A)+\Gamma_{WZ}(B) \,+\, \int d^2x \,tr\left[%
\tilde{A}\left(\partial_\tau A \partial_\sigma B - \partial_\sigma A
\partial_\tau B\right)\tilde{B}\right]\;\;.
\end{equation}

Next, we can bring results (\ref{pcm2}) and (\ref{topol}) into the
WZWN action \cite{witten,novikov}, which is described by 
\begin{equation}  \label{WZW}
{\cal S}_{WZWN}(g) = {\frac{1}{{\lambda^2}}}{\cal S}_{PCM}(g)+{\frac{n}{{4\pi%
}}} \Gamma_{WZ}(g)\;\;.
\end{equation}

\noindent We mention the appearance of an extra parameter, both in the
action and in the canonical formalism, playing the role of coupling
constant. In terms of the chiral variables, the WZWN model reads 
\begin{eqnarray}  \label{wzw2}
{\cal S}_{WZWN}\left(A,B\right)&=&\left[{\frac{1}{\lambda^2}} {\cal S}%
_{\varepsilon}(A)+{\frac{n}{4\pi}}\Gamma_{WZ}(A)\right] +\left[{\frac{1}{%
\lambda^2}}{\cal S}_{-\varepsilon}(B)+{\frac{n}{4\pi}} \Gamma_{WZ}(B)\right]
\nonumber \\
&+& \left({\frac{\varepsilon}{\lambda^2}}+{\frac{n}{4\pi}}\right) \int d^2x
tr\left[\tilde{A}\left(\partial_\tau A\partial_\sigma B -\partial_\sigma
A\partial_\tau B\right)\tilde{B}\right]\;\;.
\end{eqnarray}

\noindent We can appreciate that the separability condition is only achieved
at the critical points, as expected. But also that our choice of $%
\varepsilon $ is now dependent on which of the critical points we choose: $%
4\pi\varepsilon=- \lambda^2 n$. The result is the non-Abelian version of the
dual projection, and corresponds to the sum of two Lagrangians describing
non-Abelian chiral bosons of opposite chiralities, each one having the form
proposed by Sonnenschein \cite{sonnenschein}. We will see that a change of
the critical point automatically switches the chirality of $A$ and $B$ by
changing the sign of $\varepsilon$. Indeed, in order to obtain separability,
we must have either 
\begin{equation}  \label{sepcon1}
(i)\:\:\:\:\:\:\:{\frac{{\lambda^2n}}{{4\pi}}}=-\varepsilon = 1
\end{equation}

\noindent or 
\begin{equation}  \label{sepcon2}
(ii)\:\:\:\:\:\:\:{\frac{{\lambda^2n}}{{4\pi}}}=-\varepsilon = -1\;\;.
\end{equation}

\noindent In the first case we find the set of chiral equations as 
$$\partial_x\left(\tilde{A}\partial_+A\right)=0 \qquad \mbox{and} \qquad \partial_-\left(\tilde{B}%
\partial_xB\right)=0$$ 

\ni whose solution reads respectively 
$$A\,=\,A_-(x^-)h_A(t) \qquad \mbox{and} \qquad B\,=\,h_B(t)B_+(x^+)\;\;.$$

\ni In the second case, the chiral equations are 
$$
\partial_x\left(\tilde{A}\partial_-A\right)=0 \qquad \mbox{and} \qquad \partial_+\left(\tilde{B}
\partial_xB\right)=0
$$

\ni and the solutions read 
$$
A\,=\,A_+(x^+)h_A(t) \qquad \mbox{and} \qquad B\,=\,h_B(t)B_-(x^-)\;\;.
$$ 

\ni The arbitrary functions of time $h_A(t)$ and $h_B(t)$ 
represent, in fact, the zero modes of the solutions of the chiral equations.  
Note that in both cases the general solution for $%
g(x^+,x^-) $ is given as 
$$
g=A_\varepsilon(x^\varepsilon)h_A(t)h_B(t)B_{-%
\varepsilon}(x^{-\varepsilon})\;\;.
$$ 

\ni As in the Abelian case, the constraint $%
h_A(t)=h_B^{-1}(t)$ becomes necessary in order that $g=AB$ satisfies the
equation of motion for the WZWN model. This constraint is the only memory
left for the chiral bosons stating that they belong to the same non-chiral
field. More details about the dual projection of PCM and the coupling with
gravity can be seen in \cite{bw}.

\section{The new formulations of Hull's noton}

There are indications that a deeper understanding of such issues as string
dynamics and fractional quantum Hall effect phenomenology can be achieved by
treating the chiral sectors in a more independent way. However, coupling
chiral fields to external gauge and gravitational fields is problematic. As
we said above (it was discussed in \cite{bw}) the coupling of chiral
(Abelian) fields to external gravitational backgrounds can be achieved by
diagonalization (dual projection) of the first-order form of a covariant
scalar action. The theory reduces then to a sum of a left and a right FJ's
actions \cite{fj}, circumventing the problems caused by the lack of manifest
Lorentz invariance. 

In this section we intend to supply the literature with new formulations of
Hull's noton: the chiral WZWN model (using the dual projection), with two different 
versions of PST formulations of the chiral bosons and with the supersymmetric case. 
We hope that the use of the dual projection formalism and consequently the disclosure of a Hull's noton mode inside these models may
help to gain a new insight into the structures of these theories. The
redefinition of the fields in the first-order form of the action naturally
reveals the two-dimensional internal structure hidden into the theory.

Note that in the last subsection, what we showed was the dual projection
applied in a nonchiral model. Now the interesting touch is to see what will
be disclosed in a well defined chiral model.

\subsection{The chiral WZWN model}

We know that the action for the chiral WZWN model \cite{witten,novikov} for
a particle which moves to the left, the so-called left mover is given by 
\begin{equation}  \label{leftzero}
S\,(g)=\int\;d^2x\; tr\left(\partial_+g\:\partial_-\tilde
g\,+\,\lambda\partial_-g\:\partial_-\tilde g\right) +\Gamma_{WZ}(g)
\end{equation}
where, for convenience we write $\lambda=\lambda_{++}$ and $g \in G$ was
defined in the last section. This action can be seen as the WZWN action
immersed in a gravitational background, with a characteristic truncated
metric tensor: 
\begin{equation}  \label{leftzerograv}
S\,(g)={\frac{1}{2}}\int d^2 x \sqrt{-\eta_+}\;
\eta_+^{\mu\nu}\:tr\left(\partial_\mu g\:\partial_\nu \tilde g\right)
+\Gamma_{WZ}(g)
\end{equation}

\noindent with $\eta^+ = det(\eta^+_{\mu\nu})$ and

\begin{equation}  \label{metric+}
{\frac{1}{2}} \sqrt{-\eta_+}\: \eta_+^{\mu\nu}=\left( 
\begin{array}{cc}
0 & {{\frac{1}{2}}} \\ 
{{\frac{1}{2}}} & {\lambda}
\end{array}
\right)
\end{equation}
where $\lambda$ transforms like in the Abelian case and $g$ as a scalar.

In Lorentz coordinates we can write (\ref{leftzero}) as 
\begin{eqnarray}
S\,(g)\,=\,\int\,d^2\,x\,tr \left[ {\frac{1 }{2}}\,(1+\lambda)\dot{g}\dot{%
\tilde{g}}\, \,-\,{\frac{1 }{2}}\,(1-\lambda)\,g^{\prime}\tilde{g}%
^{\prime}\,-\,{\frac{\lambda }{2}}(\,\dot{g}\,\tilde{g}^{\prime}\,+\,g^{%
\prime}\,\dot{\tilde{g}}\,) \right] \,+\,\Gamma_{WZ}(g)
\end{eqnarray}

This action can be written in its Faddeev-Jackiw's first-order form making
the following transformation which introduces an auxiliary field ${\cal P}$, 
\[
{\frac{1}{2}}\,\dot{g}\,\dot{\tilde{g}} \rightarrow {\cal P}\,\dot{g}\,+\,{%
\frac{1 }{2}}\,{\cal P}\,g\,{\cal P}\,g 
\]
where ${\cal P}=\dot{\tilde{g}}$.

Now we can write that 
\begin{eqnarray}  \label{firstorder}
S\,(g)&=&\int\,d^2\,x\,tr \left[ (1\,+\,\lambda){\cal P}\,\dot{g}\,+\,{\frac{%
1 }{2}}\,(1+\lambda)\,{\cal P}\,g\,{\cal P}\,g \,-\,{\frac{1}{2}}%
(1-\lambda)\,g^{\prime}\,\tilde{g}^{\prime} \,-\,{\frac{\lambda }{2}}\,(\,%
\dot{g}\,\tilde{g}^{\prime}\,+\,g^{\prime}\,\dot{\tilde{g}}\,) \right] 
\nonumber \\
&+&\,\Gamma_{WZ}(g)\;\;.
\end{eqnarray}
For convenience, let us promote the following transformation of variables,

\[
{\cal P}=\frac{1}{1+\lambda}\,{\cal P}\;\;, 
\]

\noindent and substituting in (\ref{firstorder}) we have that, 
\begin{eqnarray}  \label{firstorder2}
S\,(g)\,&=&\,\int\,d^2\,x\,tr \left[ {\cal P}\,\dot{g}\,+\,{\frac{1 }{2}}\,{%
\frac{1 }{(1+\lambda)}}\,{\cal P}\,g\,{\cal P}\,g \,-\,{\frac{1}{2}}%
(1-\lambda)\,g^{\prime}\,\tilde{g}^{\prime} \,-\,{\frac{\lambda }{2}}\,(\,%
\dot{g}\,\tilde{g}^{\prime}\,+\,g^{\prime}\,\dot{\tilde{g}}\,) \right] 
\nonumber \\
&+&\,\Gamma_{WZ}(g)\;\;.
\end{eqnarray}

\noindent Making again another transformation, i.e., 
\[
{\cal P} \rightarrow {\cal P}\,+\,\lambda\,\dot{\tilde{g}} 
\]
hence, 
\begin{equation}  \label{firstorder3}
S\,(g)\,=\,\int\,d^2\,x\,tr \left[ {\cal P}\,\dot{g}\,+\,{\frac{1 }{2}}\,%
\frac{1}{1+\lambda}\,({\cal P}\,g\,{\cal P}\,g\,-\,2 \lambda {\cal P}
g^{\prime}) \,-\,{\frac{1 }{2}}{\frac{1 }{(1+\lambda)}}\,g^{\prime}\,\tilde{g%
}^{\prime}\right]  \nonumber \\
\,+\,\Gamma_{WZ}(g)\;\;.
\end{equation}

Let us redefine the fields $g$ and ${\cal P}$ through the following canonical
transformation, 
\begin{eqnarray}  \label{nove}
g\,=\,N\,h \qquad \Rightarrow \qquad
g^{\prime}\,=\,N^{\prime}\,h\,+\,N\,h^{\prime}\;\;,
\end{eqnarray}
where $N$, as before, will play the role of a Hull's nonmover field, the
noton. 

Using (\ref{novepi}) to justify a first choice for ${\cal P}$, 
let us construct it with two numerical (or not) coefficients which will be
determined in a further analysis, 
\begin{equation}
{\cal P}\,=\,a\,\tilde{h}\,\tilde{N}^{\prime }\,+\,b\,{\tilde{h}}^{\prime }\,%
\tilde{N}\;\;,  \label{noveb}
\end{equation}
but note the non-Abelian feature of this different construction. An
important observation is that the coefficients $a$ and $b$ incorporate the
roles of the separability points described in the last section. We will show
below that this form of the field ${\cal P}$ is the only possible general
form in order to promote the dual projection.

As we mentioned, it can be shown \cite{bw} that the dual projection of the
Wess-Zumino term, using the redefinition (\ref{nove}) has the form 
\begin{eqnarray}  \label{dez}
\Gamma_{WZ}(N,h) = \Gamma_{WZ}(N)\,+\,\Gamma_{WZ}(h)\,+\,\int\,d^2\,x\,tr
\left( \tilde{N}\,N^{\prime}\dot{h}\,\tilde{h}\, -\,\tilde{N}\,\dot{N}%
\,h^{\prime}\,\tilde{h}\right)\;\;.
\end{eqnarray}
We can see that the dual projection of (\ref{dez}) brings an extra term that
can not be split, so this term has to be eliminated. To perform this
elimination it is easy to see that the cast of the field ${\cal P}$ has to
have some term proportional to the extra term in (\ref{dez}). Other forms of such field
do not afford the elimination of this term, as we said before.

The canonical transformations (\ref{nove}) and (\ref{noveb}) lead us to an
action with fields taking values in the internal space. Substituting the
redefinitions (\ref{nove}), (\ref{noveb}) and (\ref{dez}) into (\ref
{firstorder3}) we can show that, 
\begin{eqnarray}  \label{onze}
& &S(N,h)\,=\, \Gamma_{WZ}\,(N)\,+\,\Gamma_{WZ}(h) \,+\, \int\,d^2\,x\,tr
\left\{ a\,\tilde{N}^\prime\,\dot{N}\,+\,b\,\tilde{h}^\prime\,\dot{h}
\,-\,(a-1)\,\tilde{N}\,\dot{N}\,\dot{h} \tilde{h} \right.  \nonumber \\
&-&\left. (b+1)\,\tilde{N}\,\dot{N}\,h^\prime \tilde{h}\, \,-\,{\frac{1 }{{%
1+\lambda}}}\left({\frac{a^2 }{2}}\,+\,\lambda a\,+\,{\frac{1 }{2}}\right)\,%
\tilde{N}^\prime\,{N}^\prime \,-\,{\frac{1 }{{1+\lambda}}}\left({\frac{b^2 }{%
2}}\,+\,\lambda b\,+\,{\frac{1 }{2}}\right)\,\tilde{h}^\prime\,{h}^\prime
\right.  \nonumber \\
&+&\left. 2\,(ab+1)\,\tilde{N}^\prime\,{N}\,h\,\tilde{h}^\prime \right\}\;\;.
\end{eqnarray}

\noindent To eliminate the extra terms we have to find $a$ and $b$ solving
the following very simple system 
\begin{equation}
a\,-\,1 \,=\, 0 \qquad \mbox{and} \qquad b\,+\,1 \,=\, 0\;\;,
\end{equation}
which solution is 
\begin{equation}
a\,=\,1 \qquad \mbox{and} \qquad b\,=\,-\,1 \qquad \Longrightarrow \qquad
a\,b\,=\,-\,1\;\;.
\end{equation}

Instead of applying directly these solutions, let us promote a short analysis making 
$a=-b=\epsilon$. With this new form for the solution, the canonical transformations are 
\begin{equation}
g\,=\,N\,h \qquad \mbox{and} \qquad p\,=\,\epsilon\,\tilde{h}\tilde{N}%
^\prime\,-\,\epsilon\,\tilde{h}^\prime\,\tilde{N}\;\;,
\end{equation}
and substituting it in (\ref{onze}) we have the action given by 
\begin{eqnarray}
S(N,h)&=&\int\,d^2\,x\,tr \left[ \epsilon\,\dot{N}\,\tilde{N}^{\prime}\,-\, 
\frac{\epsilon^2\,+\,2 \lambda \epsilon\,+\,1}{2(1+\lambda)}\,N^{\prime}\,%
\tilde{N}^{\prime} \right]\,+\,\Gamma_{WZ}\,(N)  \nonumber \\
&-&\int\,d^2\,x\,tr \left[ \epsilon\,\dot{h}\,\tilde{h}^{\prime}\,-\, \frac{%
\epsilon^2\,-\,2 \lambda \epsilon\,+\,1}{2(1+\lambda)}\,h^{\prime}\,\tilde{h}%
^{\prime} \right]\,+\,\Gamma_{WZ}\,(h) \;\;,
\end{eqnarray}
and finally we can say that 
\begin{eqnarray}  \label{notons}
S(N,h)_{\epsilon=1}&=&\int\,d^2\,x\,tr \left(\dot{N}\,\tilde{N}%
^{\prime}\,+\,N^{\prime}\,\tilde{N}^{\prime} \right)\,+\,\Gamma_{WZ}\,(N) 
\nonumber \\
&-&\int\,d^2\,x\,tr \left( \dot{h}\,\tilde{h}^{\prime}\,+\,\eta\,h^{\prime}\,%
\tilde{h}^{\prime} \right)\,+ \,\Gamma_{WZ}\,(h) \;\;,  \nonumber \\
S(N,h)_{\epsilon=-1}&=&-\int\,d^2\,x\,tr \left(\dot{N}\,\tilde{N}%
^{\prime}\,+\,\eta\,N^{\prime}\,\tilde{N}^{\prime} \right)
\,+\,\Gamma_{WZ}\,(N)  \nonumber \\
&-&\int\,d^2\,x\,tr \left(\dot{h}\,\tilde{h}^{\prime}\,+\,h^{\prime}\,\tilde{%
h}^{\prime} \right)\,+ \,\Gamma_{WZ}\,(h) \;\;,
\end{eqnarray}
where 
\[
\eta\,=\,\frac{1-\lambda}{1+\lambda}\;\;. 
\]
Since we are working in a $D=2$ dimensions, i.e., $0$-form $(D=2(2p+1))$,
each reduced phase space carries the representation for half the number of
degrees of freedom of the original system. This is a feature different from $%
D=4p$, where the duality symmetric systems maintains the phase space
structure intact \cite{baw2}. For the interested read, an analysis of the
WZWN duality groups is depicted in \cite{aal}.

We see in (\ref{notons}) that in accordance with the value of $\epsilon$, $N$ and $%
h$ change roles. For $\epsilon=1$, $N$ is the chiral field and $h$ is the
non-Abelian Hull's noton.   For $\epsilon=-1$, vice versa. This shows a
certain dependence of the dual projection on a certain parameter, but it is
immaterial since what we want to show is the presence, in the chiral WZWN,
of a chiral mode and mainly the presence of a non-Abelian Hull's noton.

This result corroborates the one found in \cite{wotzasek2}. However, there,
the noton was found using the soldering technique \cite{solda} of two
non-Abelian chiral bosons, showing a destructive interference. Our work
shows a different way to obtain the same result. Besides, we explain the
result in \cite{wotzasek2} in a clearer way since now we can see both chiral
particles and realize what is happening. As the WZWN model has a chiral mode
and an algebraic mode, in the fusion (soldering) of opposite non-Abelian
chiral models, the opposite chiral particles interfere destructively
disappearing from the spectrum. Only the Hull noton survives, which as
having no dynamics, does not interact destructively.  
In other words, we show that the Hull noton was already there, on the chiral WZWN model and consequently it is not just a product of soldering both opposite models as can be understood from \cite{wotzasek2}.

\subsection{The PST self-dual formulation}

It can be shown \cite{pst2} that one of the possibilities to recover the
manifest Lorentz invariance of the FJ model, as in the case of Maxwell theory, is introducing an
unit-norm time-like auxiliary vector field $u_{m}\,(x)$ in the
FJ action, and to write the action in the form 
\begin{equation}
S\,=\,\int d^{2}x\,\left( \,\partial _{+}\phi \,\partial _{-}\phi
\,-\,u^{m}{\cal F}_{m}u^{n}{\cal F}_{n}\,\right)  \label{pst}
\end{equation}
where ${\cal F}_{m}=\partial _{m}\phi \,-\,\epsilon _{mn}\partial ^{n}\phi $%
. Because of its properties, $u_{m}$ contains only one independent
component. It was proved in \cite{pst2} that (\ref{pst}) reduces to the Siegel action \cite
{siegel} and also that $u_m$ is the gradient of a scalar.  It will be used below.

In Minkowski space we can write the Lagrangian density of (\ref{pst}) as 
\begin{equation}
{\cal L}\,=\,\dot{\phi}^{2}\,-\,{\phi ^{\prime }}^{2}\,-\,(u^{0}\,+%
\,u^{1})^{2}\,(\dot{\phi}\,+\,\phi ^{\prime })^{2}
\end{equation}
and following the procedure to promote the dual projection in first-order actions 
we have, after a little algebra, that 
\begin{equation}
{\cal L}\,=\,\pi \,\dot{\phi}\,-\,\frac{\left[ \pi
\,+\,2(u^{0}+u^{1})^{2}\phi ^{\prime }\right] ^{2}}{4\left[
1-(u^{0}+u^{1})^{2}\right] }\,-\,[1\,+\,(u^{0}+u^{1})^{2}]\,{\phi ^{\prime }}%
^{2}  \label{pst2}
\end{equation}
where $\pi$ is the canonical momentum.

Now, as before, we can use the convenient canonical transformations,  
\begin{equation}
\phi \,=\,\pm {\frac{1}{\sqrt{2}}}\,(\rho \,+\,\sigma )\qquad \mbox{and}%
\qquad \pi \,=\,\pm \sqrt{2}\,(\rho ^{\prime }\,-\,\sigma ^{\prime })\;\;.
\end{equation}
where one of these fields will represent the FJ particle and the other, of course, 
the noton, as we will see.

Substituting these transformations in (\ref{pst2}) and after an algebraic
work we have an action in terms of $\sigma$ and $\rho$, which take values in
the internal space, 
\begin{equation}
{\cal L}\,=\,\left( -\,\sigma ^{\prime }\,\dot{\sigma}\,-\,{\sigma ^{\prime }%
}^{2}\right) \,+\,\left( \rho ^{\prime }\,\dot{\rho}\,-\,\eta _{1}\,{\rho
^{\prime }}^{2}\right)  \label{44}
\end{equation}
where 
\[
\eta _{1}\,=\,\frac{1+(u^{0}+u^{1})^{2}}{1-(u^{0}+u^{1})^{2}}\;\;. 
\]
As we can easily see the $\sigma $ field describes a FJ chiral boson that now, at first sight, carries the dynamics of the system and the $Z_{2}$ duality group. However, in \cite
{baw2} it was shown that in all even dimensions, both $SO(2)$ and $Z_{2}$
duality symmetric groups could co-exist. As we have analyzed in the anterior
section, the $\rho $ field is responsible for the algebra \cite{pst2} of the
system. Hence, $\rho $ is the PST Hull's noton formulation.

The other possibility (see \cite{pst2}), which is more appropriate from the quantum point of view, is to construct a Lorentz covariant action like the following 
\begin{equation}
S\,=\,\int d^{2}x\,\left( \partial _{+}\phi \,\partial _{-}\phi \,+\,{%
\frac{1}{u^{2}}}\,u^{m}{\cal F}_{m}u^{n}{\cal F}_{n}\,-\,\epsilon
^{mn}\,u_{m}\partial _{n}B\right)  \label{pst3}
\end{equation}
where $B(x)$ is an auxiliary scalar field. After eliminating $B$ through the field equations and taking the values for $u_{m}$, the Lagrangian density of the above action can be
written as 
\begin{equation}
{\cal L}\,=\,\partial _{+}\phi \,\partial _{-}\phi \,-\,{\frac{\partial
_{+}\hat{\varphi}}{\partial _{-}\hat{\varphi}}}\,(\partial _{-}\phi
)^{2}\;\;,  \label{pst4}
\end{equation}
where $\hat{\varphi}$ is an another auxiliary field that helps in the
solution of the equation for $u_{m}$, i.e., $u_{m}(x)=\p_{m}\hat{\varphi}(x)$.  Hence, $u_{m}(x)$ can be seen as the gradient of a scalar $\hat{\varphi}(x)$.

In this case it is easy to see that the equation (\ref{pst4}) has the same
form as the Siegel chiral boson, where 
\[
\lambda \quad \rightarrow \quad {\frac{\partial_{+} \hat{\varphi} }{%
\partial_{-} \hat{\varphi}}} 
\]

\noindent and with a straightforward association we already have the
solution given in \cite{aw}

\begin{equation}
{\cal L}\,=\,\rho ^{\prime }\,\dot{\rho}\,-\,{\rho ^{\prime }}%
^{2}\,-\,\sigma ^{\prime }\,\dot{\sigma}\,-\,\eta _{2}\,{\sigma ^{\prime }}%
^{2}\;\;,
\end{equation}
where we have again $\sigma $ as the noton and 
\[
\eta _{2}\,=\,\frac{1\,-\,{\frac{\partial _{+}\,\hat{\varphi}}{\partial
_{-}\,\hat{\varphi}}}}{1\,+\,{\frac{\partial _{+}\,\hat{\varphi}}{\partial
_{-}\,\hat{\varphi}}}}\;\;. 
\]
In the face of this result we can make an analogous analysis for this second
possibility of Lorentz invariance restoration as we made it for the first one
and stress that the dynamics/$(Z_{2}+SO(2))$ duality groups and algebra
group are carried by $\rho$ and $\sigma $ respectively. We see clearly that
Hull's noton is present whatever the way we formulate the Siegel chiral
boson.

\subsection{The supersymmetric formulation}

The superfield generalization of (\ref{pst3}) is 
\begin{equation}  \label{super}
S\,=\,\int d^2 x d\theta^+ \,\left( D_+ \Phi\,\partial_{-}\Phi \,-\,{\frac{%
D_{+} \Lambda }{\partial_{-} \Lambda}}\,(\partial_{-}\,\Phi)^2 \right)
\end{equation}
where we are considering the bosonic superfields 
\[
\Phi\,(x^{-},x^{+},\theta^+)\,=\,\phi(x)\,+\,i\,\theta^+\,\psi_+ (x) 
\]
and 
\[
\Lambda\,(x^{-},x^{+},\theta^+)\,=\,\hat{\varphi}(x)\,+\,i\,\theta^+\,%
\chi_+ (x) \;\;, 
\]
which obey the conventional transformation laws under global shifts 
\[
\delta \theta^+\,=\,\epsilon^+ \;,\qquad \delta
x^{+}\,=\,i\theta^+\,\epsilon^+ \qquad \mbox{and} \qquad \delta
x^{-}\,=\,0 
\]
in $n=(1,0)$ flat superspace and where 
\begin{eqnarray}
D_+\,&=&\,{\frac{\partial }{\partial\theta^+}}\,+\,i\,\theta^+\,\partial_{+}
\nonumber \\
D^2_+\,&=&\,i\,\partial_{+}
\end{eqnarray}
is the supercovariant derivative.

After making the relevant substitutions described above in (\ref{super}) we
can write 
\begin{equation}
S\,=\,\int d^2 x d\theta^+ \left\{ \,{\frac{\partial \Phi }{\partial \theta^+%
}}\,\partial_{-}\,\Phi \,-\,\,{\frac{\partial \Lambda }{\partial \theta^+}}%
\,\,{\frac{(\partial_{-} \Phi)^2 }{\partial_{-} \Lambda}} \,+\,
\,i\,\theta^+\,\left[ \partial_{+}\Phi\,\partial_{-}\Phi \,-\,\,{\frac{%
\partial_{+} \Lambda }{\partial_{-} \Lambda}}\,(\partial_{-}\Phi)^2 %
\right] \right\}\;\;.
\end{equation}
Using the well known properties of the supersymmetric integrals we finally
have that, 
\begin{equation}  \label{super2}
{\cal L}\,=\, \partial_{+}\Phi\,\partial_{-}\Phi \,-\,{\frac{\partial_{+}
\Lambda }{\partial_{-} \Lambda}}\,(\partial_{-}\Phi)^2 \;\;,
\end{equation}
where a hidden global complex number is of no consequence here since it does
not affect the equations of motion and may be ignored. With this form, we
can make in (\ref{super2}) the same association that we made in (\ref
{pst4}). So, we can write, considering that now we are in a superspace that, 
\begin{equation}
{\cal L}\,=\, \Sigma^\prime\,\dot{\Sigma}\,-\,{\Sigma^\prime}%
^2\,-\,\Gamma^\prime\,\dot{\Gamma} \,-\,\eta_3\,{\Gamma^\prime}^2\;\;,
\end{equation}
where 
\[
\eta_3\,= \,\frac{1\,-\,{\frac{\partial_{+}\,\Lambda }{\partial_{-}\,%
\Lambda}}} {1\,+\,{\frac{\partial_{+}\,\Lambda }{\partial_{-}\,\Lambda}}}
\;\;. 
\]
The superfield $\Sigma$ represents the super-FJ chiral boson in the superspace
formulation and $\Gamma$ is the supersymmetric Hull's noton, the supernoton%
\footnote{%
A very interesting study of the duality groups in supersymmetric models can
be seen in \cite{rv}.}.

\section{Conclusion}

We have diagonalized the non-Abelian chiral field, disclosing a two
dimension internal structure for some compact Lie group $G$. This result was
expected based on the results found in \cite{aw} where it was shown that
notons, being not dynamicals, couples to the gravitational backgrounds but
not to the electromagnetic field. Therefore, if a
gauge coupling is introduced before dual projection, it will be completely
decoupled by the dual projection procedure. Since the analysis was always
effected for first-order systems, an equivalence between the Lagrangian and
Hamiltonian approaches permitted us to use the concept of canonical
transformations. In other words we can say that the dual projection demanded
a change of variables which was, in the phase space, a canonical
transformation.

In \cite{pst2} it was proposed two ways to restore the manifest Lorentz
invariance of the FJ theory. The first was introducing an unit-norm time-like auxiliary
vector field in the FJ action and constructing an action
equivalent to the Siegel model. The other is constructing an action in the
light of the Maxwell manifestly Lorentz invariant duality symmetric action
proposed in \cite{pst2}, which is shown to be also Siegel's equivalent on
the mass shell. We show that in both cases the noton is present. Our last
result concerns the superfield generalization of the above action and a
supersymmetric formulation of the noton, the supernoton has been constructed.

As a straightforward perspective we can diagonalize superiors orders of
chiral $p$-forms and the investigation of the dual projection of superfield
supergravity generalization of the duality symmetric models is in course.


\section{Acknowledgments}

The author would like to thank the Abdus Salam International Centre for
Theoretical Physics and acknowledge the help of the High Energy Section and
the Science Institute of the Federal University of Itajub\'{a} (Minas Gerais, Brazil) where
part of this work was made. And professor C. Wotzasek for valuable
discussions. This work was partially done at the High Energy section within
the framework of the guest-scientist Scheme of the Abdus Salam ICTP.

\end{document}